\documentclass[a4paper, amsfonts, amssymb, amsmath, reprint, showkeys, nofootinbib, twoside,superscriptaddress,aps,prl]{revtex4-2}
\usepackage[pdftex, pdftitle={Article}, pdfauthor={Filipe R.N.C. Maia}]{hyperref} 
\usepackage{graphicx}

\begin{document}
\title{3D-Printed Sheet Jet for Stable Megahertz Liquid Sample Delivery at X-ray Free Electron Lasers}

\author{Patrick E. Konold}
\author{Tong You}
\affiliation{Laboratory of Molecular Biophysics, Institute for Cell and Molecular Biology, Uppsala University, Box 596, 75124 Uppsala,  Sweden}

\author{Johan Bielecki}
\author{Joana Valerio}
\author{Marco Kloos}
\affiliation{European XFEL, Holzkoppel 4, 22869, Schenefeld, Germany}

\author{Daniel Westphal}
\affiliation{Laboratory of Molecular Biophysics, Institute for Cell and Molecular Biology, Uppsala University, Box 596, 75124 Uppsala,  Sweden}

\author{Alfredo Bellisario}
\author{Tej Varma}
\author{August Wolter}
\affiliation{Laboratory of Molecular Biophysics, Institute for Cell and Molecular Biology, Uppsala University, Box 596, 75124 Uppsala,  Sweden}

\author{Jayanath C. P.Koliyadu}
\author{Faisal H.M. Koua}
\author{Romain Letrun}
\author{Adam Round}
\author{Tokushi Sato}
\affiliation{European XFEL, Holzkoppel 4, 22869, Schenefeld, Germany}

\author{Petra Mésźaros}
\author{Leonardo Monrroy}
\author{Jennifer Mutisya}
\affiliation{Department of Chemistry - BMC, Uppsala University, Box 576, 75123, Uppsala, Sweden}

\author{Szabolcs Bódizs}
\author{Taru Larkiala}
\author{Amke Nimmrich}
\affiliation{Department of Chemistry and Molecular Biology, University of Gothenburg, Gothenburg, Sweden}

\author{Roberto Alvarez}
\affiliation{Department of Physics, Arizona State University, 550 E. Tyler Dr., Tempe, AZ, 85287, USA}

\author{Richard Bean}
\affiliation{European XFEL, Holzkoppel 4, 22869, Schenefeld, Germany}
\author{Tomas Ekeberg}
\affiliation{Laboratory of Molecular Biophysics, Institute for Cell and Molecular Biology, Uppsala University, Box 596, 75124 Uppsala, Sweden}
\author{Richard A. Kirian}
\affiliation{Department of Physics, Arizona State University, 550 E. Tyler Dr., Tempe, AZ, 85287, USA}
\author{Sebastian Westenhoff}
\affiliation{Department of Chemistry - BMC, Uppsala University, Box 576, 75123, Uppsala, Sweden}
\affiliation{Department of Chemistry and Molecular Biology, University of Gothenburg, Gothenburg, Sweden}

\author{Filipe R. N. C. Maia}
\email[E-mail: ]{filipe.maia@icm.uu.se}
\affiliation{Laboratory of Molecular Biophysics, Institute for Cell and Molecular Biology, Uppsala University, Box 596, 75124 Uppsala,  Sweden}
\affiliation{Lawrence Berkeley National Laboratory, Berkeley, CA, 94720, USA}

\date{\today} 

\begin{abstract}
X-ray Free Electron Lasers (XFELs) can probe chemical and biological reactions as they unfold with unprecedented spatial and temporal resolution. A principal challenge in this pursuit involves the delivery of samples to the X-ray interaction point in such a way that produces data of the highest possible quality and with maximal efficiency. This is hampered by intrinsic constraints posed by the light source and operation within a beamline environment. For liquid samples, the solution typically involves some form of high-speed liquid jet, capable of keeping up with the rate of X-ray pulses. However, conventional jets are not ideal because of radiation-induced explosions of the jet, as well as their cylindrical geometry combined with the X-ray pointing instability of many beamlines causes the interaction volume to differ for every pulse. 
This complicates data analysis and contributes to measurement errors. An alternative geometry is a liquid sheet jet which, with its constant thickness over large areas, eliminates the X-ray pointing related problems. Since liquid sheets can be made very thin, the radiation-induced explosion is reduced, boosting their stability. They are especially attractive for experiments which benefit from small interaction volumes such as fluctuation X-ray scattering and several types of spectroscopy. Although they have seen increasing use for soft X-ray applications in recent years, there has not yet been wide-scale adoption at XFELs. Here, we demonstrate liquid sheet jet sample injection at the European XFEL SPB/SFX nano focus beamline. We evaluate several aspects of its performance relative to a conventional liquid jet including thickness profile, stability, and radiation-induced explosion dynamics at high repetition rates. The sheet jet exhibits superior performance across these critical experimental parameters. Its minute thickness also suggests ultrafast single-particle solution scattering is a possibility.
\end{abstract}

\maketitle
\section{Introduction}
X-ray Free Electron Lasers (XFELs) have revolutionized the field of molecular imaging and spectroscopy by offering ultrashort X-ray pulses with very high peak brilliance and spatial coherence. Such instruments have offered an unparalleled glimpse of molecular machinery with atomic-scale resolution \cite{chapman_2011, kraus_2018, bergmann_2021}. Despite the remarkable development of XFELs over the past years, several obstacles remain to further push the boundaries of methodological development to yield more refined molecular information.

Liquid sample injection is an essential element of XFEL research that impacts a wide range of measurement applications. This spans powerful biological imaging methods such as serial femtosecond crystallography (SFX) and solution scattering that are direct structural probes of condensed phase molecular reactions, as well several types of spectroscopies with a range of applications. However, given the unique characteristics of XFEL light sources, care must be taken to achieve proper sample delivery. Inevitable radiation-induced damage resulting from the extremely high peak intensity necessitates that measurements are conducted serially and samples must be delivered under continuous flow \cite{vakili_2022}. This is especially challenging at high-repetition-rate beamlines, such as the European XFEL (EuXFEL), which is capable of producing MHz pulse trains. Additionally, large background fluctuations arising from turbulent liquid flow, excessive sample consumption, vacuum compatibility, and ongoing dilemmas concerning efficient and reproducible nozzle fabrication remain an ongoing hindrance for interrogation of liquid specimens.

Currently, the gas dynamic virtual nozzle (GDVN) represents the most common means of liquid sample injection at XFELs. These devices employ a sheath of high velocity gas, which encapsulates and accelerates liquid within a central channel \cite{gancalvo_1998, deponte_2008}. This effect produces a high velocity cylindrical liquid jet emerging from the nozzle, followed by rapid disintegration into a droplet stream. Such nozzles, now routinely produced by 3D-printing, yield jets of varying diameter (from sub-µm to 100s of µm) with flow rates approaching µLs per minute \cite{vakili_2022, knoka_2020, nazari_2020, nelson_2016}. These characteristics help to reduce background from the surrounding solvent and avoid excessive sample consumption. Equally important is their velocity, which must be on the order of 10 m/s to outpace the radiation-induced jet explosion and subsequent shockwave generated by preceding pulses \cite{stan_2016, wiedorn_2018a}.

Despite the wide utility of GDVNs, there are several undesirable aspects of their operation that warrant further improvement. For example, their curved geometry makes it difficult to obtain stable interaction volumes with the X-ray laser and the chaotic breakup of the jet into droplets may produce unwanted scattering \cite{eggers_2008}. Moreover, experimental nuisances such as inconsistent jetting behavior and susceptibility to clogging are additional hindrances for XFEL applications that typically require extended data collection periods spanning several days.

An emerging means of solution phase sample delivery is the liquid sheet jet. In their original form, liquid sheets were most commonly generated by oblique collision of opposing laminar jets emitted from independent nozzles \cite{taylor_1960}. At intermediate flow rates, the balance of inertial forces and intrinsic properties of the liquid yields a chain of mutually orthogonal sheets \cite{bush_2004}. Various experimental and modeling efforts have revealed that the primary sheet section exhibits µm scale thickness with an extremely flat surface profile and smooth flow behavior that is stable over long periods \cite{menzi_2020, choo_2002, sanjay_2017}. Alternative nozzle designs have been recently achieved through lithography and 3D-printing that enable liquid sheet formation within a microfluidic template \cite{ha_2018, galinis_2017}. One particular variant utilizes gas acceleration, akin to GDVNs, where opposing gas channels collide with a central liquid channel \cite{koralek_2018}. This configuration produces analogous fluid chains as described above, but with dramatically thinner sheet sections, approaching 10s of nanometers, and a $\sim$10x lower flow rate. Such a nanofluidic medium with laminar flow behavior offers the enticing prospect of dramatically improving liquid sample injection applied to difficult solution-phase imaging techniques. One such example is the application of fluctuation X-ray scattering which relies on quantifying subtle intensity correlations and has so far only been demonstrated on large virus particles \cite{kurta_2017, pande_2018}. Reducing the number of particles in the interaction region would bring the signal into the range best suited to current detectors without decreasing the signal to noise ratio, which may enable the application of this powerful imaging method to smaller biomolecules. 

Liquid sheet jets have found increasing use as a means of sample delivery in vacuum for soft X-ray spectroscopy, electron diffraction and also as a medium for high harmonic generation and high intensity laser plasma investigations \cite{luu_2018, ekimova_2015, smith_2020, george_2019, wiedorn_2018, nunes_2020, yang_2021, fondell_2017}. Their use at XFEL sources has been far less prevalent, likely due to the high liquid and gas loads that complicate vacuum operation. Recently, Hoffman and coworkers reported on liquid sheet jet injection with a hard X-ray XFEL source where they observed sheet jet explosion upon exposure to a nanofocus X-ray beam at 120 Hz \cite{hoffman_2022a}. Their approach utilized impinging jet nozzles which produced sheet thickness on the order of several micrometers and flow rates of mL/min. This paper builds on this previous account and demonstrates liquid sheet jet injection at a high repetition rate hard X-ray XFEL source, the Single Particles, Clusters, and Biomolecules and Serial Femtosecond Crystallography (SPB/SFX) beamline at the EuXFEL. Moreover, we introduce a 3D-printed gas-accelerated nozzle design that can be rapidly and reproducibly fabricated, which enables efficient prototyping to expand potential experimental applications. The results illustrate key facets of their performance including stability, thickness distribution, velocity, and radiation-induced breakup dynamics at repetition rates up to 1.13 MHz. These findings set the stage for broader adoption of liquid sheet jet sample delivery at XFELs and have the potential to enhance experimental precision for solution phase X-ray experimentation. 

\section{Experimental Methods}

\subsection{3D-Printing of Liquid Sheet Nozzles}
The nozzles were produced by two-photon polymerization (TPP) using the NanoOne 3D-printing system (UpNano). Nozzle design was performed with the Solidworks computer-aided design (CAD) program (Dassault Systèmes). The resulting STL files were loaded in the instrument control software (Think3D) for optimization of printer parameters. TPP was initiated with a 800 nm laser (85 mW) focused with a 10x objective. The printing was carried out on a 10 x 10 x 5.5 mm glass substrate (silanized) in vat mode, where the glass surface was submerged in liquid resin (UpPhoto). Following printing, the completed nozzles were placed in a beaker and soaked in propylene glycol monomethyl ether acetate (PGMEA) for 1-2 days while gently stirring to remove residual unpolymerized resin. Upon sufficient development, the nozzles were washed with isopropanol and dried for future use.

\subsection{Nozzle Assembly}
Fused silica capillaries (Polymicro 0.360 mm OD / 0.150 mm ID) were used to deliver pressurized liquid and gas to the nozzles during operation. These were ground flat on the inlet side using a polishing wheel, washed in an ultrasonic bath, and dried prior to use. For assembly of the nozzle, the capillaries were manually inserted into the inlet ports under a microscope in a vertical orientation. Next, a small drop of 5 minute epoxy glue (Loctite Universal Power Epoxy) was placed slightly above the inlet ports, which gradually wicked along the capillary to yield a uniform coverage at the nozzle interface. The assembled nozzle was allowed to cure overnight before experimental use.

\subsection{Liquid Sample Injection at EuXFEL SPB/SFX}

The printed sheet jet nozzles were mounted to the standard liquid injector rod provided by the EuXFEL Sample Environment group. The rod was assembled as follows: A 1/8 inch OD stainless steel tube was first glued to the capillaries approximately 5 mm above the nozzle. The tube was fed through 10-32 PEEK fitting (Idex) which was fastened into a stainless steel nozzle adaptor. The capillaries were then fed through the entire length of the rod and the end piece was screwed into the tip

Liquid and gas were delivered to the nozzle as previously described \cite{vakili_2022}. In short, liquid reservoirs were connected to the nozzle inlets via PEEK tubing (Idex, 0.250 µm ID). Multiple sample reservoirs were connected in parallel to facilitate fast sample switching executed with a high speed electronic valve (Rheodyne). Liquid flow was regulated using an HPLC pump (Shimadzu LC-20AD), while helium gas flow was regulated with an electronic pressure regulator (Proportion Air GP1). Liquid and gas flow rates were monitored with in-line flow meters (Sensirion and Bronckhorst respectively). 

Alignment of the nozzle tip with respect to the interaction region was carried out by manipulating the position of the injector rod using motorized stages. This placement was aided by visualization with the side-view microscope camera illuminated with the EuXFEL femtosecond laser coupled into the sample chamber via a fibre bundle laser synchronized with the X-ray pulse \cite{koliyadu_2022, palmer_2019}.

\subsection{SPB/SFX Beamline Configuration}

The data were collected at the SPB/SFX instrument of the EuXFEL in September 2022, under the proposal p3046 \cite{mancuso_2019}. The EuXFEL produced bunch trains at 10 Hz with intratrain pulse repetition rates between 141 kHz and 1.13 MHz. The photon energy was 8000 eV, approximately 1.55 Å. From previous measurements, the focal spot was estimated at around 300 x 300 nm. The energy of every X-ray pulse was measured by a gas monitor detector upstream and averaged around 2 mJ. With this beamline configuration and photon energy the beamline transmission between the gas monitor detector and the interaction region is estimated to be 65\%. The AGIPD 1M detector was 0.331 m downstream from the interaction region \cite{allahgholi_2019}. The experiment was monitored online with Hummingbird \cite{daurer_2016}.

\subsection{Sheet Jet Thickness Calculation}

The thickness of the sheet jet was estimated by scaling the experimental data to the calculated solution scattering pattern \cite{cromer_1968} of a GROMACS molecular dynamics simulation \cite{abraham_2015} of the bulk liquid sample \cite{vanderspoel_2018}, taking into account the pulse energy on the sample, as measured by the X-ray gas monitor detector using the beamline transmission of 65\% (see Figure S3). The scaling factor of the fit was then used to estimate the sheet jet thickness. The same procedure was applied for GDVN thickness determination using water as the sample liquid. The determined GDVN thickness matched the expected value given its flow rates \cite{vakili_2022} .

\section{Results and Discussion}

\subsection{Nozzle Design and Fabrication}

3D-printing by TPP enables precise and efficient fabrication of microscopic structures. These advantages were exploited for the production of our liquid sheet jet nozzles. Several 3D-printed nozzles have been previously demonstrated for liquid jet generation, including GDVNs and mix-and-inject applications \cite{vakili_2022}. Throughout the design optimization process, we consulted these previous accounts with the specific objective of balancing structural stability and channel expression together with overall printing and operational efficiency. 

\begin{figure}
\includegraphics[width=\columnwidth,keepaspectratio]{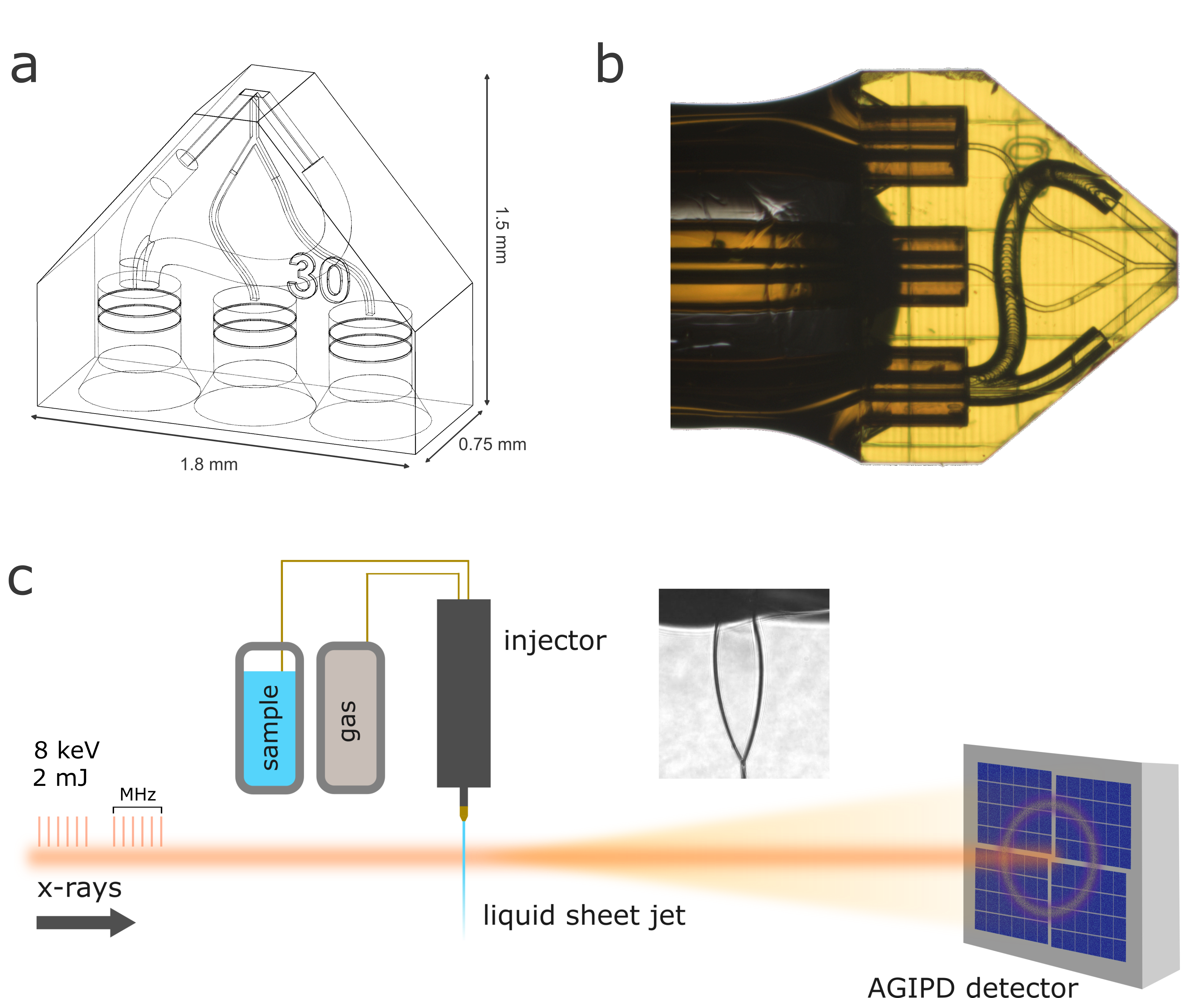}
\caption{(a) CAD drawing of 3D-printed sheet jet nozzle design. (b) Image of assembled sheet jet nozzle (c) Experimental setup at the EuXFEL SPB/SFX nanofocus beamline.}
\end{figure}

The channel geometry follows from previous iterations of gas-accelerated sheet jet nozzles, where two opposing gas channels surround a central liquid channel (Figure 1a). The liquid channel dimensions (30 x 30 µm) were chosen to mitigate clogging, while restricting the overall flow rate to minimize sample consumption and maintain vacuum compatibility. The two gas channels originated from a single inlet, which was split within the printed body of the chip. Utilizing this common gas inlet freed up additional space to accommodate two liquid channels which merged 250 µm above the tip of the nozzle. This geometry serves multiple functions in the context of maintaining consistent flow behavior within a beamline setting. For example, the second liquid channel exists as a means to extend device operation in the event of clogging. Moreover, the additional inlet might be used for introduction of a second liquid such as for mix-and-inject, on-chip droplet generation, and multiphase flow applications. 

\subsection{Liquid Sheet Jet Operation at the EuXFEL SPB/SFX Beamline}

Liquid sheet jet sample injection was carried out at the EuXFEL SPB/SFX nanofocus beamline. The experimental configuration is depicted in Figure 1c. The nozzle was fitted to the standard liquid injector rod and aligned to 45 degrees with respect to both the incoming X-rays and side view microscope for best visualization of the liquid sheet jet behavior. A snapshot of this configuration is shown in the inset of Figure 1c. The liquid and gas were delivered to the nozzle using fused silica capillaries pressurized with an HPLC pump. Liquid and gas flow rates (Q$_{L}$ and Q$_{G}$) of 75-115 µL/min and 10 mg/min for the sheet jet and 23 µL/min and 21 mg/min for the GDVN were used in these measurements. It is known that sheet jets can be generated over a wide range of liquid and gas flow rates, which may strongly influence their performance. The operating conditions here were selected to produce the most stable jetting behavior while maintaining vacuum compatibility. Isopropanol was chosen as the sample liquid for this measurement to avoid downtime due to aqueous sample freezing.

Several facets of the sheet jet operation were considered in order to evaluate its performance compared to conventional cylindrical liquid jets. First, jet stability was investigated by monitoring the integrated scattered X-ray response while focusing through the center of the primary sheet section and normalized by the incoming pulse intensity as measured by the X-ray gas monitor (XGM). This analysis was carried out on varying timescales to capture jet behavior over time as well as with different repetition rates. Figure 2a illustrates its performance for all shots compared to an analogous GDVN measurement run over a 5 minute period. Overall, the sheet jet exhibited far more consistent behavior and lower deviations on all timescales, while the GDVN displayed a slightly skewed distribution towards lower intensities. Moreover, binning this curve in 50 pulse intervals revealed a prominent 0.5 Hz resonance in the GDVN response that is absent in the sheet jet. This noise pattern may be explained when considering a couple of known factors. X-ray beam pointing fluctuations in the horizontal plane of 2-3 focal spots are typical at the SPB/SFX beamline. Given that the beam is focused on the center of the cylindrical GDVN jet ($\sim$3 µm diameter), such a displacement significantly changes the interaction volume. As such, this can lead to intermittent stochastic dropouts in the scattered intensity as observed within the lower envelope of GDVN response. 

\begin{figure}
\includegraphics[width=\columnwidth,keepaspectratio]{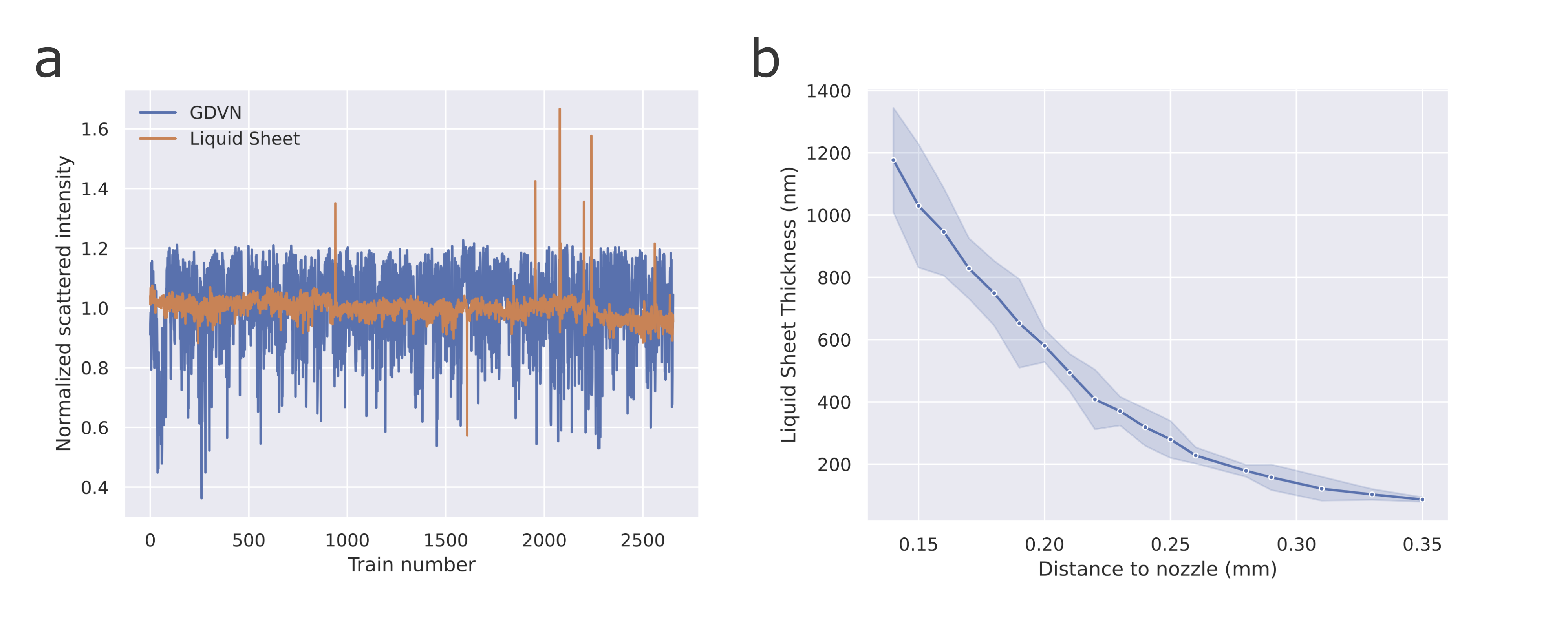}
\caption{(a) The integrated AGIPD detector response normalized by incoming X-ray intensity for all trains over a representative 5 minute measurement window for the liquid sheet jet and GDVN. (b) Thickness dependence of the primary sheet section while scanning vertically away from the nozzle tip towards the lower rim. The shaded region reflects a measurement error of 3 standard deviations.}
\end{figure}

The oscillatory behavior may require a slightly different explanation. The liquid is driven using a dual piston HPLC pump set to a fixed Q$_{L}$, while flow pulsation of $\sim$10\% is known for this pumping configuration. In this experiment, much smaller deviations in Q$_{L}$ of 1-2\% were recorded on in-line flow meters and no statistical correlation was found between the observed fluctuations and Q$_{L}$. Therefore, we suspect beam pointing fluctuations to be the dominant source of deviations observed here. Overall, the sheet jet exhibited 4-fold lower background fluctuations compared to the GDVN (3 versus 13\% standard deviation), which was representative for all repetition rates tested up to 564 kHz.

Vertical and horizontal scans were carried out to investigate trends in liquid thickness across the primary sheet section. The nanofocus X-ray beam is particularly well-suited for this task given its high spatial precision. In order to retrieve absolute thickness values, \textit{ab initio} scattering curves were calculated from MD simulations of the bulk liquid. The modeled curves were then scaled to the experimental data to yield position-dependent thickness. The results of this analysis for a vertical scan are shown in Figure 2b and the trend is in strong agreement with previous characterizations of similar nozzle variants using optical interferometry and mid-IR absorption \cite{koralek_2018}. An analogous scan in the horizontal dimension revealed very little thickness variation suggesting the primary section is extremely flat (Figure S2). This is in contrast to prior accounts of impinging jet liquid sheets, which exhibit significant curvature \cite{galinis_2017}. The thickness curves also showed no dependence with X-ray repetition rate.

The structure of the primary flat section presents intriguing possibilities for potential X-ray experimentation. The thickness values measured here represent about a 30-fold reduction in path length compared to a typical GDVN ($\sim$micrometers). Such a nanoscopic sample medium may achieve a dramatic reduction in background levels for X-ray scattering and spectroscopic investigations and allow for investigation of samples over a wide range of concentrations, approaching the single molecule scale. Moreover, its wedge-like structure enables one to quickly vary sample thickness by simple translation of the injector rod which may be useful in situations with a distinct trade-off between signal and background. The remarkable flatness in the horizontal dimension may also help overcome longstanding difficulties with conventional cylindrical liquid jets. For example, a curved sample medium complicates calculation of electron take-off angles for X-ray photoelectron spectroscopy as well as distorts pump laser focal properties in transient absorption applications. Finally, the large usable area of the liquid sheet vastly reduces sensitivity to beamline pointing fluctuations. 
  
Radiation-induced explosion due to the high peak intensity X-ray pulses is an important consideration when evaluating the behavior of liquid jets and offers valuable insight with respect to their potential experimental utility \cite{stan_2016}. Explosion of the liquid sheet jet was captured by the side view microscope within the experimental chamber and snapshots at select time delays are presented in Figure 3. The beam position is indicated by the plasma spot generated from the nanofocused X-rays, roughly centered within the flat section of the liquid sheet. Upon arrival of the pulse, an elliptical vacancy formed originating at the focal point which rapidly expanded and propagated along the direction of liquid flow until eventually exiting the lower rim of the flat section. This closely resembles the behavior observed with thicker impinging sheet jets at a lower repetition rate XFEL source \cite{hoffman_2022a}. A second bubble formed upon arrival of the subsequent X-ray pulse and followed a trajectory similar to the initial one without perturbing the X-ray focal spot. At long times, these vacancies converged towards the lower section of the liquid jet and culminated in a fraying pattern that spanned the remaining field of view. Variation of the focal position drastically altered the jet explosion dynamics. For example, focusing on the rim section led to much more complex behavior, while positioning at the thinnest point led to a less disruptive breakup presumably due to exposure of differing liquid volumes (Movie S1). X-ray repetition rate dependence of this phenomenon was also investigated. Similar jet explosion dynamics were observed at 282 and 564 kHz (Movie S2), however the breakup became much more intense at 1.13 MHz suggesting the sheet did not fully regenerate between consecutive pulses within a train at this rate. 

\begin{figure}
\includegraphics[width=\columnwidth,keepaspectratio]{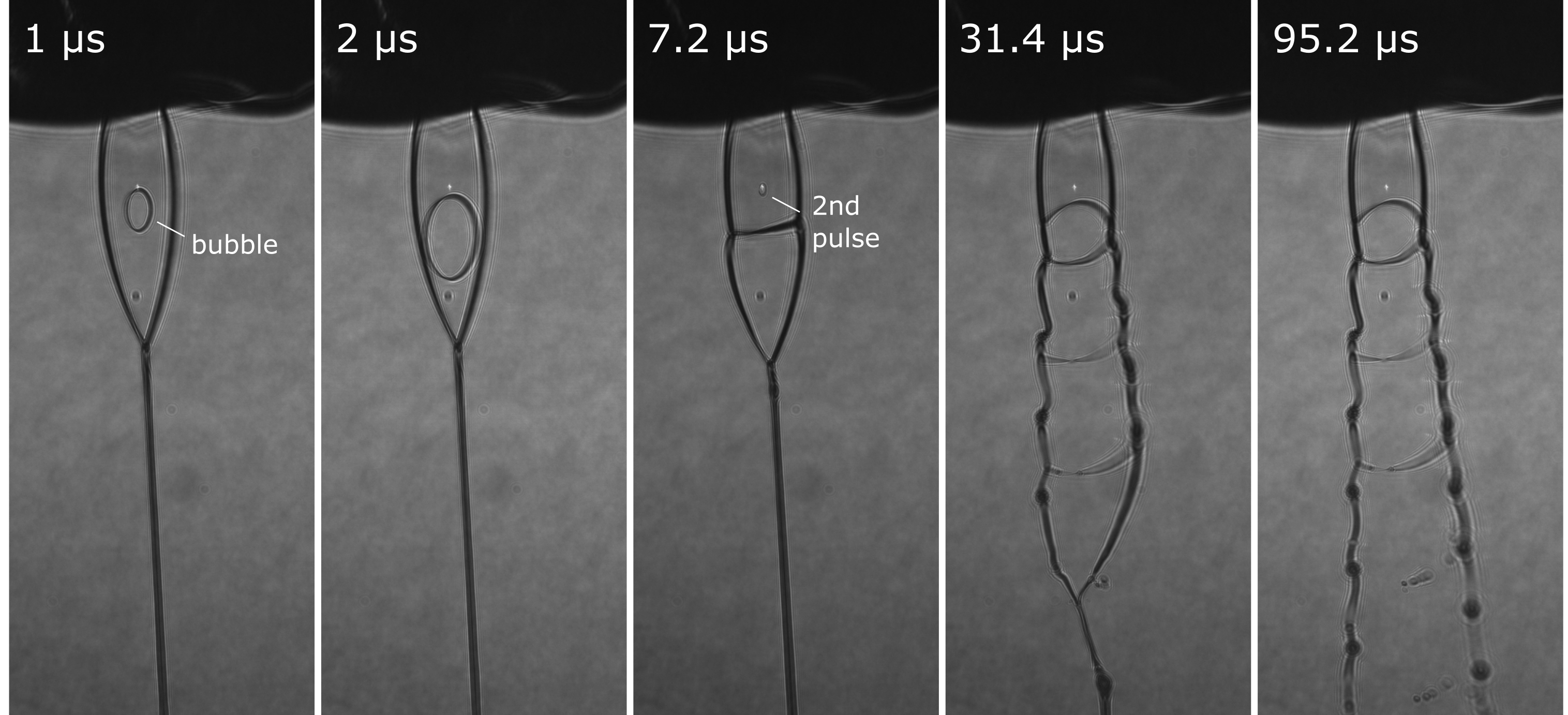}
\caption{Stroboscopic visualization of sheet jet explosion upon exposure to the nanofocus XFEL beam (141 KHz) at various time points as captured by the side microscope camera within the experimental chamber. The fluid velocity within the liquid sheet was estimated by following the movement of the bubble over time. Note that the flat section of the jet is oriented at 45 degrees with respect to the camera.}
\end{figure}

The movement of this vacancy informs on the liquid velocity within the flat section of the sheet jet, which was extracted by using frames captured by the side microscope camera accounting for the 10x objective magnification. The leading and trailing expansion fronts exhibited different velocities. The trailing edge maintained a roughly constant velocity of 11 m/s throughout its propagation, while the leading edge showed some acceleration, initially moving at 60 m/s near the focal region and increasing to 110 m/s as it approached the lower rim. The trailing edge velocity is similar to previous values reported for impinging sheet jets \cite{dombrowski_1954, hoffman_2022a}. The shape of this vacancy may also reflect a parabolic velocity profile within the flat section as has been previously predicted \cite{choo_2002}.

The above results demonstrate the striking potential of liquid sheet jet sample injection for X-ray measurements at XFEL beamlines. It exhibited higher stability and shorter sample path length over a much larger target area compared to the current standard liquid jet (GDVN). Moreover, the radiation-induced explosion did not perturb detection of the diffracted X-rays for repetition rates up to 564 kHz. Such features should benefit multiple X-ray experimental methods such as time-resolved solution scattering, fluctuation X-ray scattering \cite{kirian_2011} and several spectroscopy modalities. Thus, if properly implemented, this platform has the potential to transform solution-phase sample injection at XFEL light sources and allow new experiments such as single-particle solution scattering. 

Despite these encouraging results, several operational nuances must be addressed prior to their broader deployment. A few relevant factors are described below: 

\subsection{Sample consumption} Average liquid flow rates on the order of 100 µL/min are typical for gas-accelerated sheet jet nozzles. A large fraction of this flow is contained within the annular rings that surround the flat sheet sections. An unfortunate consequence of this geometry is that the vast majority of liquid volume passes through the interaction region without being interrogated leading to excessive sample waste. A couple strategies might be employed to overcome this issue. Scaling down the nozzle geometry would reduce the overall liquid flux. However, this would also increase clogging susceptibility - a chronic problem for microfluidic flow applications which scales quickly for channel dimensions of less than 30 µm. Moreover, the lower overall target area might be problematic for applications with large focal spots. An alternative approach is to confine the sample specimen within a surrounding carrier liquid, akin to double flow focusing GDVN nozzles \cite{oberthuer_2017}. A similar idea was recently demonstrated by Hoffman et al. where layered aqueous/non-aqueous heterostructures were contained within the flat liquid sheet section \cite{hoffman_2022}. However, in this case, a thin aqueous region was bounded by two thicker nonaqueous layers. Such a configuration would effectively dilute the overall sample response since the X-rays must traverse through all liquid layers. An alternative approach with sample focusing in the orthogonal plane would potentially resolve this issue. Additional sample consumption savings might be achieved through introduction of a segmented flow scheme.

\subsubsection{Vacuum startup} Seamless startup in vacuum is an essential feature required to avoid extended measurement interruptions during beamtime. The nozzles employed in this study presented a unique challenge in this regard. Long capillaries (\textgreater2 m) are required to deliver liquid and gas to the nozzle, which dictates that high line pressures are necessary to achieve adequate flow rates at the nozzle tip. Given that HPLC pumps build pressure slowly (over several seconds), it can be troublesome to establish reliable jetting while starting under vacuum. To overcome this challenge, the pump was first primed against a plug to 1000 psi and then quickly switched to the nozzle line using an electronic valve. Although this “burst” method drastically improved vacuum startup, it may be less desirable over the long run given that the high pressures required place extra stress on the tubing fittings within the line. Other strategies might be considered. First, one could substitute a large segment of the small ID capillary (150 µm) for larger ID PEEK tubing (\textgreater250 µm). This would drastically reduce the line resistance and lessen the pressure needed to establish reliable jetting. A second alternative would be to utilize a different means of driving liquid flow. For example, pressurizing the sample reservoir directly with finely regulated air would enable much faster line pressurization and help facilitate more robust vacuum startup.

\subsubsection{Vacuum compatibility} The elevated liquid loads required to run liquid sheet jets create an exceptional challenge for prolonged operation within a high vacuum environment. Several schemes involving differential pumping, cryo-trapping, and heated catching devices have been previously employed to facilitate liquid jet operation in vacuum \cite{hoffman_2022a, galinis_2017}. In the current study, a stainless steel shroud was used to isolate the volume surrounding the nozzle tip and a catcher was used to contain liquid at the bottom of the chamber. Main chamber pressure stabilized in the range of 10$^{-4}$  mbar and was suitable for prolonged operation. While this configuration was successful for running isopropanol in this instance, the use of aqueous samples might place different constraints on liquid sheet jet use in vacuum (e.g. icing issues). Moreover, given the tight tolerances posed by this system, developing a standard solution broadly compatible across many beamlines and facilities might be difficult given the wide variety of vacuum configurations that exist.

\subsubsection{External Meniscus} In the current nozzle design, the liquid meniscus resides along the exit orifice on the outer surface of the nozzle tip. During vacuum operation, small droplets formed on this surface (visible in the frames shown in Figure 3) and persisted over time. Given that this effect was not observed during operation at atmospheric pressure, we attribute it to a broadening of the meniscus that occurs under vacuum. When the liquid reaches the nozzle edge, a strong wicking action pulls it across the entire outer surface forming a droplet. Once this droplet reaches some critical size, it gets pulled further upward into the above capillaries, rapidly destabilizes and detaches. This process occurred every few minutes and repeated several times during the measurement. Such droplet formation, while mainly a nuisance during our measurements, is undesirable for long duration jet operation. One possible solution to minimize this effect is to recess the meniscus within the nozzle body as is typical for GDVNs. In this way, the surrounding gas might act to buffer the liquid away from the outer surface of the nozzle and prevent droplet formation. It is worth noting that this effect primarily occurs with alcohol solutions. In separate tests while running aqueous sample in vacuum, residual liquid on the nozzle surface quickly froze and sublimated given the difference in vapor pressure.

\section{Conclusion}

Investigation of liquid phase sample specimens using high precision X-ray techniques presents a unique challenge at XFELs. A critical element of these applications involves the choice of sample delivery, which often dictates the success of a given experiment. High speed liquid jets currently represent the standard means of liquid sample injection, however they suffer from several drawbacks that limit their experimental utility. Liquid sheet jets have previously shown promise to overcome these issues, but they had not been tested at high-repetition rate XFELs. 

In this study we demonstrated liquid sheet jet sample injection at the EuXFEL SPB/SFX beamline. We used a 3D-printed gas-accelerated nozzle design to produce sheet jet thicknesses below 100 nm resulting in a significantly more stable scattering signal compared to a conventional GDVN. Furthermore, the radiation-induced explosion was found to not perturb data collection for repetition rates approaching MHz. This account also serves as a practical guide for implementation of sheet jet injection systems where vacuum compatibility is an ongoing challenge given their elevated liquid flow rates. These results demonstrate the great potential of sheet jets for high repetition rate liquid sample injection and set the stage for wider adoption at beamline facilities. With sheet jet thicknesses comparable with the ice layer in a cryogenic electron microscopy sample they also suggest the tantalizing possibility of doing ultrafast single-particle solution scattering.

Experimental data collected in this study has been deposited at the Coherent X-ray Imaging Data Bank \cite{maia_2012} (\href{https://www.cxidb.org/id/218.html}{cxidb.org/id/218.html}). The data analysis scripts and nozzle CAD files used herein can be found at \href{https://github.com/FilipeMaia/3D_MHz_liquid_sheet}{github.com/FilipeMaia/3D\_MHz\_liquid\_sheet}.



\begin{acknowledgments}

We acknowledge European XFEL in Schenefeld, Germany, for provision of X-ray free-electron laser beam time at the SPB/SFX  instrument and would like to thank the staff for their assistance. We acknowledge the use of the European XFEL biological sample preparation laboratory, enabled by the XBI User Consortium. We acknowledge valuable discussions with David van der Spoel. The results of the work were obtained using Maxwell computational resources operated at Deutsches Elektronen-Synchrotron (DESY), Hamburg, Germany. We also acknowledge Myfab Uppsala for providing facilities and experimental support. Myfab is funded by the Swedish Research Council (2019-00207) as a national research infrastructure. This work was supported by the Röntgen-Ångström Cluster (2019-06092); the Swedish Research Council (2017-05336 and 2018-00234); the Swedish Foundation for Strategic Research (ITM17-0455); Carl Tryggers Stifetelse (CTS 19-227); and the National Science Foundation (Awards DBI-1231306, DBI-1943448, MCB-1817862).

\end{acknowledgments}


\bibliography{exported-references_no_url} 

\end{document}